\documentclass[a4paper]{article}
\usepackage{epsfig,amsmath,amsfonts,amssymb,setspace,multirow,textcomp,xspace,
url}
\usepackage[T1]{fontenc}
\makeatletter
\renewcommand{\@oddhead}{\textit{Advances in Astronomy and Space Physics} \hfil}
\renewcommand{\@evenfoot}{\hfil \thepage \hfil}
\renewcommand{\@oddfoot}{\hfil \thepage \hfil}
\makeatother

\renewenvironment{thebibliography}[1]{\begin{oldthebibliography}{#1}\setlength{
\parskip}{0ex}\setlength{\itemsep}{0ex}}{\end{oldthebibliography}}
\newcommand{\xmm}{\textit{XMM-Newton}\xspace}

\begin{document}
\fontsize{11}{11}\selectfont %

\title{Creation of 2-5\:keV and 5-10\:keV sky maps using \xmm\ data}
\author{\textsl{D.\,O.~Savchenko$^{1}$, D.\,A.~Iakubovskyi$^{1,\,2}$}}
\date{\vspace*{-6ex}}
\maketitle
\begin{center} {\small $^{1}$Bogolyubov Institute of Theoretical Physics,
Metrologichna str. 14-b, 03680, Kyiv, Ukraine\\ $^{2}$National University
``Kyiv-Mohyla Academy'', Skovorody str. 2, 04070, Kyiv, Ukraine\\
{\tt dsavchenko@bitp.kiev.ua}}
\end{center}

\begin{abstract}
Sky maps are powerful visualisation tools for quicklook analysis of extended
sources. The latest sky map in soft 
X-rays (0.1-2.4~keV) has been created in 90ies using ROSAT data. By analyzing
publically available data from \xmm\ 
X-ray mission we constructed new sky maps in two energy bands -- 2-5\:keV and
5-10\:keV, complementary to ROSAT data, covering about 1\% of all sky, and included 
them to our web-based tool \url{http://skyview.virgoua.org}. \\[1ex]
\end{abstract}

\section*{\sc Introduction}

\indent \indent Usually, astronomers deal with catalogues of \emph{point}
sources. However, if the source is 
\emph{extended} (i.e. its size is comparable or even bigger than the point
spread function of the instrument),
more sophisticated method of scientific data visualisation is needed. The most
common method of such 
visualisation is building \emph{sky maps} -- specially processed series of
2-dimensional images in different energy bands.
An example of such map for X-ray astronomy is all-sky map in 0.1-2.4~keV band
made by ROSAT X-ray 
satellite~\cite{Snowden:95,Voges:99} observations. This all-sky map also exists
as interactive web-tool~\cite{RASS}.

After the end of ROSAT mission, several missions in keV range have
been operating. These missions have covered 
a minor part of the sky (not more than several \%) but with much better
sensitivity and wider energy range compared 
with ROSAT. In this paper, we present the interactive maps in 2-5 and 5-10~keV
range. For these maps, we use publicly 
available observations by MOS cameras~\cite{Turner:00} of \xmm~\cite{Jansen:01}
X-ray mission. Special attention was 
paid to handle with most important background components, including soft proton
flares and quiescent particle background, 
see~\cite{BGtable} for detailed properties of \xmm\ background. The obtained map
is included to 
website of Virtual Roentgen and Gamma Observatory in
Ukraine~\footnote{\url{http://skyview.virgoua.org}}.

\begin{figure}[!h]
\centering
\begin{minipage}[t]{.45\linewidth}
\centering
\epsfig{file = 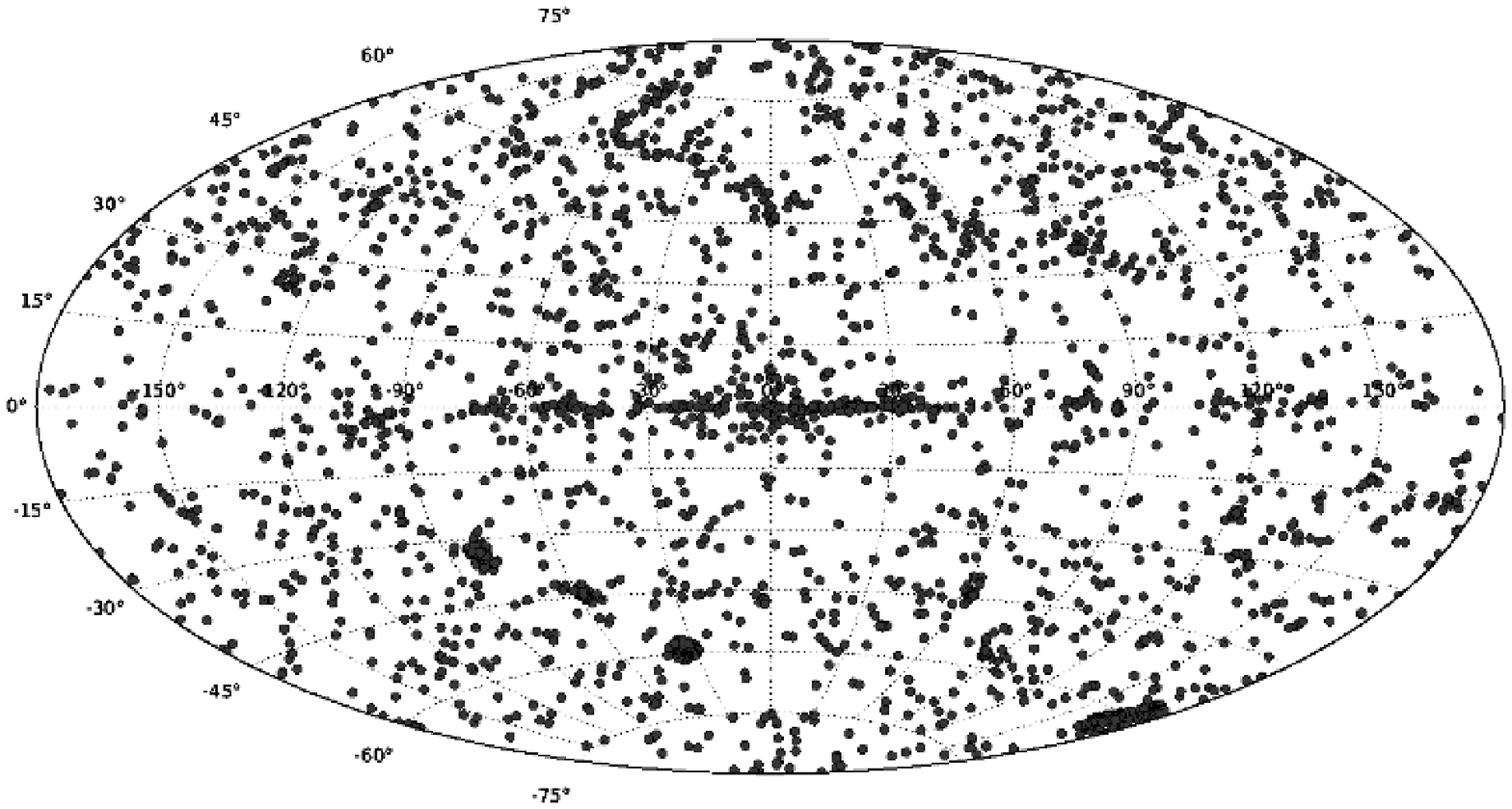,width = 1.05\linewidth}
\caption{Positions (in galactic coordinates) of \xmm\ observations used in our
analysis. The field-of-views of \xmm\ 
observations are given in natural values, so one can easily recognize the zones
observed by \xmm\
covering about 1\% of all sky.}\label{fig:obs-positions}
\end{minipage}
\hfill
\begin{minipage}[t]{.45\linewidth}
\centering
\epsfig{file = 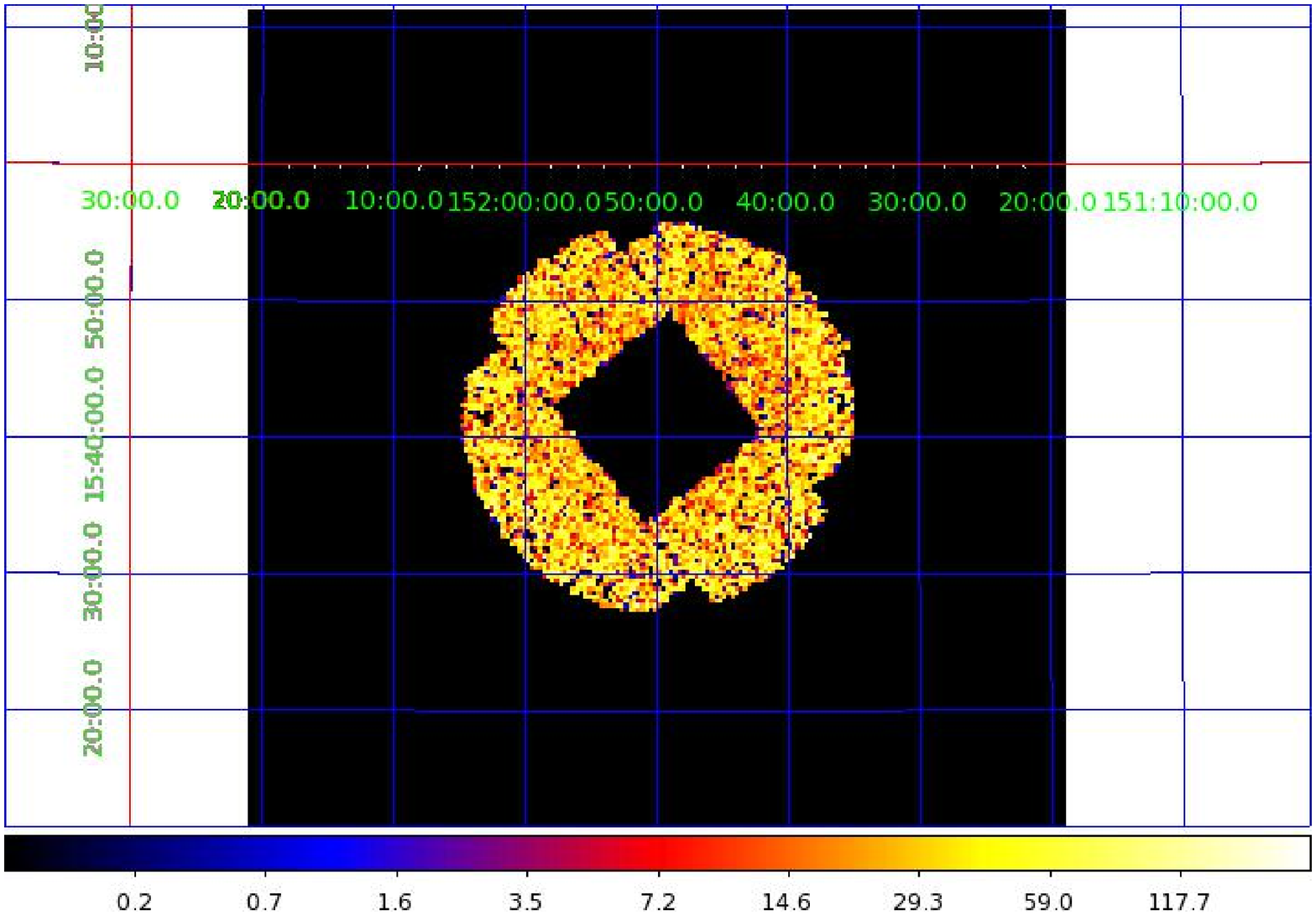,width = 0.95\linewidth}
\caption{An example of very bright \emph{point} source observation -- polar BY~Cam in
2-5~keV. The units are in cts/s/deg$^2$.
The position of BY~Cam coincides with central CCD of MOS instruments. Because
this point source is very bright 
($\gtrsim 1$~mCrab~\protect\cite{Baumgartner:12}), it was not observed in usual imaging mode and therefore
appears as a ``gap'' in the 
sky map.}\label{fig:bycam-2-5-200x200}
\end{minipage}
\end{figure}

\begin{figure}[!h]
\centering
\begin{minipage}[t]{.45\linewidth}
\centering
\epsfig{file = 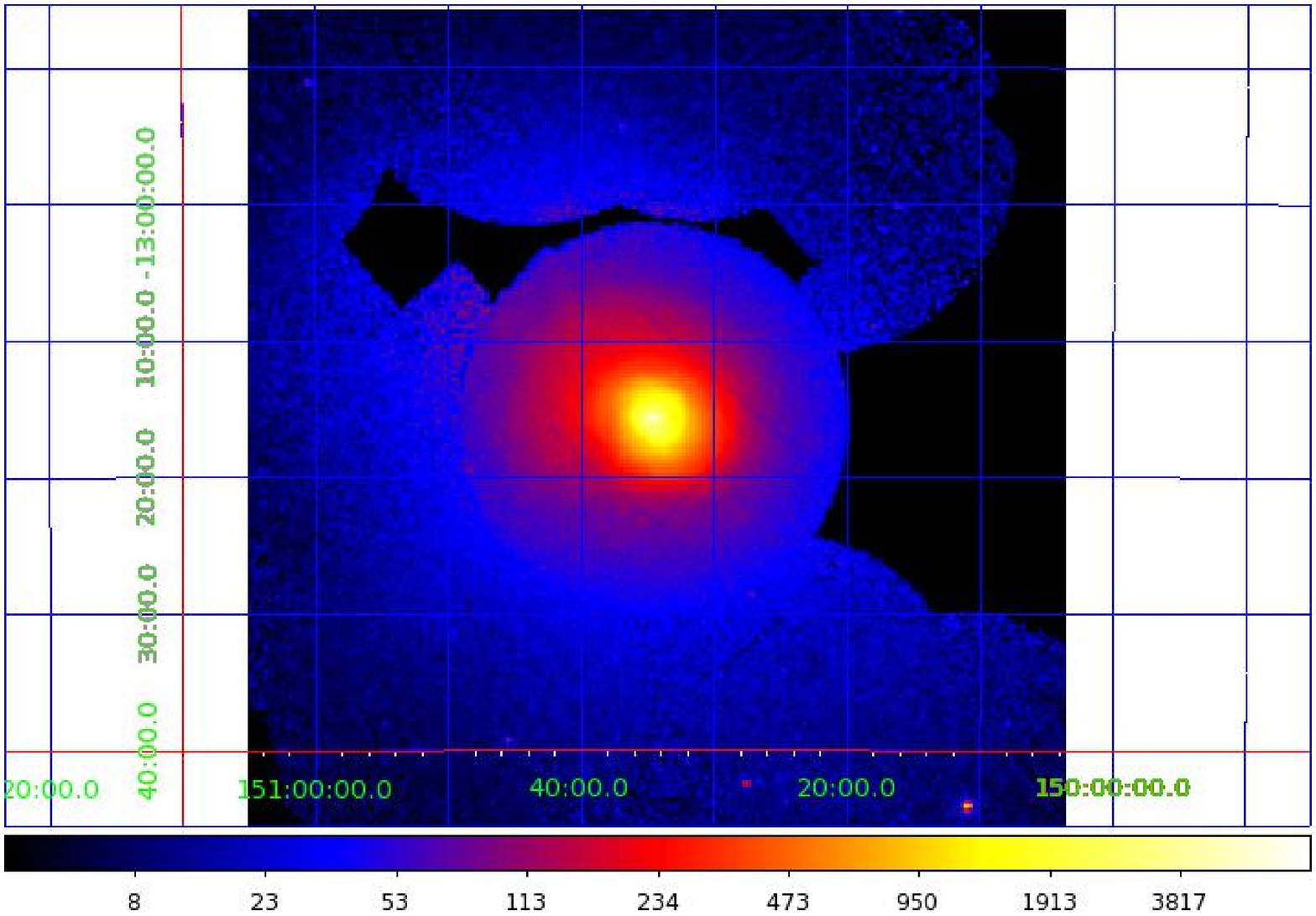,width = 0.95\linewidth}
\caption{A 2-5~keV image of 1 square degree around Perseus cluster of galaxies.
The units are in cts/s/deg$^2$.}
\label{fig:gc-2-5-200x200}
\end{minipage}
\hfill
\begin{minipage}[t]{.45\linewidth}
\centering
\epsfig{file = 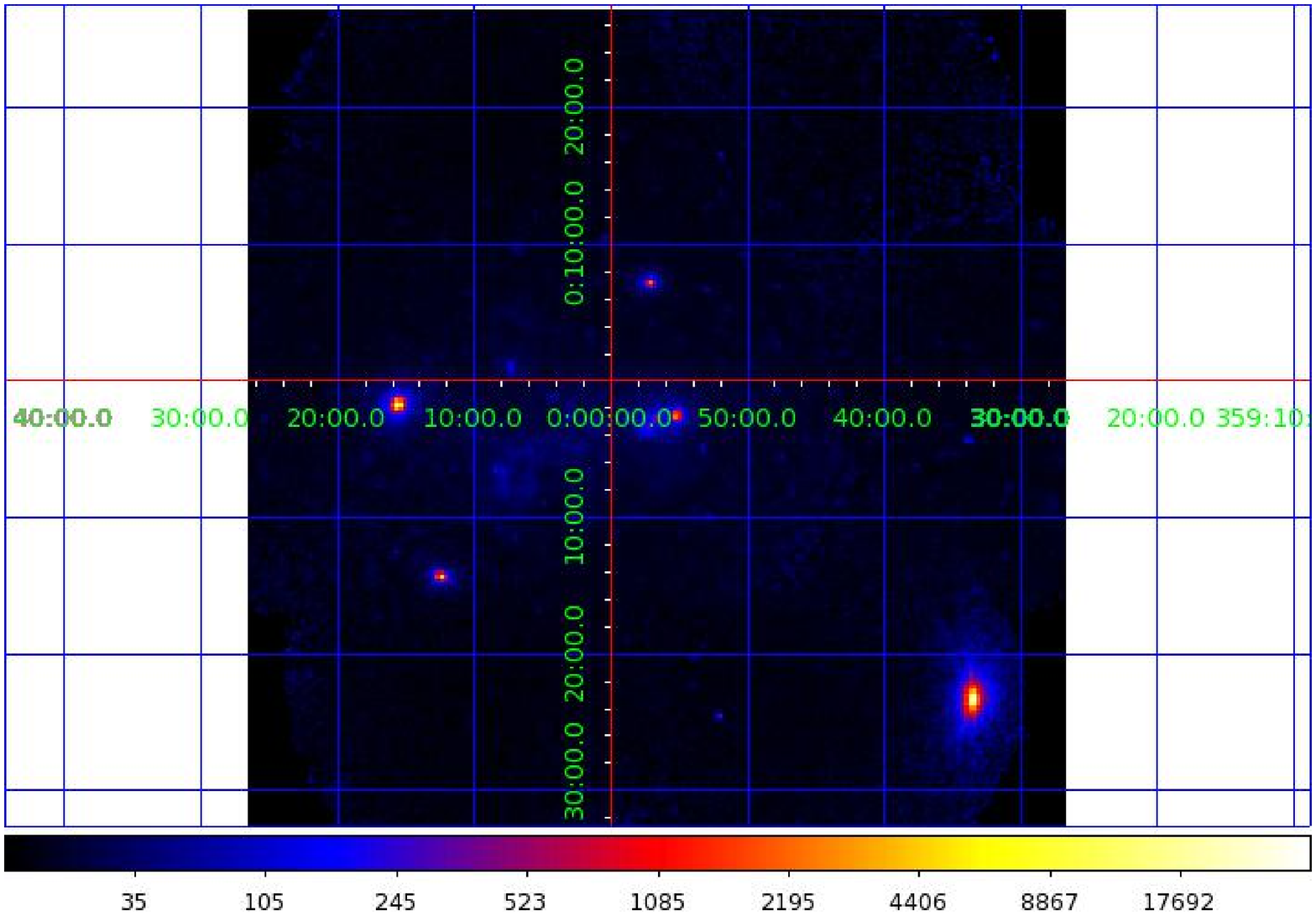,width = 0.95\linewidth}
\caption{The same as at the left Figure but for Milky Way centre in 5-10~keV
range.}
\label{fig:perseus-5-10-200x200}
\end{minipage}
\end{figure}

\begin{table}
 \centering
 \caption{General properties of MOS observations used in our
analysis.}\label{tab:obs-properties}
 \vspace*{1ex}
 \begin{tabular}{ccc}
  \hline
  Camera                 & MOS1 & MOS2 \\
  \hline
  No. of observations    & 3942 & 4022 \\
  No. of data files      & 4029 & 4104 \\
  Cleaned exposure, Ms   & 77.9 & 81.5 \\
 \hline 
 \end{tabular}
\end{table}

\section*{\sc Methods}

\indent \indent For constructing sky map, we first downloaded all publically
available (on July~1, 2013) 
observation data files for MOS~\cite{Turner:00} cameras of \xmm\ X-ray
observatory~\cite{Jansen:01} 
available on the HEASARC data archive~\cite{HEASARC}. These data files were
processed using \emph{Extended Sources Analysis Software} (ESAS)
package~\cite{ESAS-cookbook,Kuntz:08} specially developed for analysis of extended sources at the 
NASA/GSFC \xmm\ Guest Observer Facility~\cite{NASA-GOF} 
in cooperation with the \xmm\ Science Operation Centre~\cite{XMM-SOC} and the
\xmm\ Background Working Group~\cite{XMM-BG}.
It is publically available as part of \xmm\ Science
Analysis System (SAS) v.13.5.0.
The methodology of ESAS software is based on detailed modeling and/or subtraction 
of various background components (see~\cite{BGtable} for complete list) experienced by MOS and PN cameras 
on-board \xmm\ cosmic mission using the ``first principles'' as much as possible.
To model instrumental background, ESAS software relies on filter-wheel-closed data and the 
data from the unexposed corners of archived observations, rather than ``blank sky'' data (contaminated 
by unknown level by different variable background components) used by a number of 
other methods. This is essential for analysis of very faint sky regions 
(e.g. galaxy cluster outskirts) dominated by the background (rather than the source) emission.
The obtained data products -- filtered event\footnote{Here, 
an ``event'' is a result of instantaneous positive detection in one or several adjacent CCD 
pixels. 
 Single photon hitting the CCD may produce substantial signal in adjacent pixels causing so-called
\emph{multuple} (e.g. double, triple, quadruple) events.
The standard selection procedure used in our analysis takes into account single, double,
triple and quadruple events for MOS cameras. According to~\cite{Lumb:02} the procedure based 
on analysis of event patterns allows to reject of about 99\% of events caused by high-energy 
($\sim 100$~MeV) cosmic rays thus significantly reducing the amount of data telemetry.} 
lists, images, lightcurves and spectra -- are produced in 
FITS~\cite{Greisen:02,Calabretta:02} format for user-defined regions within \xmm\ field-of-view.

Our data reduction is started from production of filtered event lists using ESAS script \texttt{mos-filter}.
This script effectively removes
time intervals affected by highly variable background component -- soft proton
flares, see~\cite{BGtable,Kuntz:08}. We used the standard filters and cuts provided by ESAS software.
For example, we selected single, double, triple and quadruple events (described by event
\texttt{PATTERN} $<= 12$) of highest quality (described by \texttt{FLAG} $== 0$)\footnote{Such selection based
on \texttt{FLAG} keyword excludes events out of instrument FoV, near CCD corners and ``hot pixels'' etc. It is
generally recommended~\protect\cite{XMM-UHB,XMM-USG} to select \texttt{FLAG} $== 0$ for high-quality spectral 
analysis.}.
Main parameters for obtained event lists are shown in Table~\ref{tab:obs-properties}.
The leftover MOS event lists were processed by ESAS scripts \texttt{mos-spectra}
and \texttt{mos\_back} giving
observed and modelled quiscent particle background spectra, exposure maps, count
images for selected energy ranges and 
modelled particle background count images.
The resulting images and exposure maps of individual observations are then
combined by ESAS scripts \texttt{merge\_comp\_xmm} and \texttt{bin\_image\_merge} into 
count-rate images of sky regions with size $22^\circ\times22^\circ$ and minimal pixel size 
$2.5''\times2.5''$.
Point sources are not excluded, although very bright point sources 
observed with timing mode (such as BY~Cam, see Fig.~\ref{fig:bycam-2-5-200x200}) 
haven't been processed by ESAS and therefore do not appear in our map. For the sky map, 
we chose two energy ranges --
2-5\:keV and 5-10\:keV -- motivated by their
\begin{itemize}
 \item negligible contamination by remaining Solar Wind Charge Exchange
background component, see~\cite{BGtable,Kuntz:08}
for details;
 \item complementarity to existing ROSAT all-sky map~\cite{RASS} in 0.1-2.4~keV.
\end{itemize}

For sky map visualisation, we used the standard NASA \texttt{skyview.jar} tool~\cite{skyview}.
This tool selects appropriate images overlapping with given sky region and samples them to the given pixel size. 
The Sutherland-Hodgman clipping algorithm was used to resample images. This method treats the output pixel 
grid as a window over the input images grid and integrates the flux within each output pixel exactly. 
The output image can be produced in given sky coordinates and projection.
The obtained images in \texttt{FITS}~\cite{Greisen:02,Calabretta:02} format are available for 
quick look and can be directly downloaded from~\url{http://skyview.virgoua.org}.

\section*{\sc Results}

We constructed sky maps in 2-5\:keV and 5-10\:keV bands using $\sim$4000
publically available observations of MOS cameras on-board \xmm\ X-ray cosmic mission. 
Positions of given observations and their basic properties are shown in
Fig.~\ref{fig:obs-positions}
and Table~\ref{tab:obs-properties}, respectively.
The produced maps are cleaned from variable soft proton component and instrumental background with
the help of standard analysis for extended sources -- ESAS software~\cite{ESAS-cookbook} --
and included to web-interface of Virtual Roentgen and
Gamma-Ray Observatory in Ukraine, \url{http://skyview.virgoua.org}, see
Figs~\ref{fig:gc-2-5-200x200} 
and~\ref{fig:perseus-5-10-200x200} as examples. The obtained maps cover about 1\% of all sky,
see Fig.~\ref{fig:obs-positions} for details.
They are complementary to existing ROSAT all-sky map 
in soft X-rays (0.1-2.4~keV) as well as usual X-ray catalogues of point sources.

\section*{\sc acknowledgement}
\indent \indent We thank Yuri Izotov, Vladimir Savchenko, Igor Telezhinsky, Ievgen Vovk
and the anonymous Referee for their comments and suggestions.
This work was supported by part by the Program of Cosmic Research of the National Academy of 
Sciences of Ukraine, the State Programme of Implementation of Grid Technology in Ukraine 
and the grant of President of Ukraine for young scientists.

%


\let\jnlstyle=\rm\def\jref#1{{\jnlstyle#1}}\def\aj{\jref{AJ}}
  \def\araa{\jref{ARA\&A}} \def\apj{\jref{ApJ}\ } \def\apjl{\jref{ApJ}\ }
  \def\apjs{\jref{ApJS}} \def\ao{\jref{Appl.~Opt.}} \def\apss{\jref{Ap\&SS}}
  \def\aap{\jref{A\&A}} \def\aapr{\jref{A\&A~Rev.}} \def\aaps{\jref{A\&AS}}
  \def\azh{\jref{AZh}} \def\baas{\jref{BAAS}} \def\jrasc{\jref{JRASC}}
  \def\memras{\jref{MmRAS}} \def\mnras{\jref{MNRAS}\ }
  \def\pra{\jref{Phys.~Rev.~A}\ } \def\prb{\jref{Phys.~Rev.~B}\ }
  \def\prc{\jref{Phys.~Rev.~C}\ } \def\prd{\jref{Phys.~Rev.~D}\ }
  \def\pre{\jref{Phys.~Rev.~E}} \def\prl{\jref{Phys.~Rev.~Lett.}}
  \def\pasp{\jref{PASP}} \def\pasj{\jref{PASJ}} \def\qjras{\jref{QJRAS}}
  \def\skytel{\jref{S\&T}} \def\solphys{\jref{Sol.~Phys.}}
  \def\sovast{\jref{Soviet~Ast.}} \def\ssr{\jref{Space~Sci.~Rev.}}
  \def\zap{\jref{ZAp}} \def\nat{\jref{Nature}\ } \def\iaucirc{\jref{IAU~Circ.}}
  \def\aplett{\jref{Astrophys.~Lett.}}
  \def\apspr{\jref{Astrophys.~Space~Phys.~Res.}}
  \def\bain{\jref{Bull.~Astron.~Inst.~Netherlands}}
  \def\fcp{\jref{Fund.~Cosmic~Phys.}} \def\gca{\jref{Geochim.~Cosmochim.~Acta}}
  \def\grl{\jref{Geophys.~Res.~Lett.}} \def\jcp{\jref{J.~Chem.~Phys.}}
  \def\jgr{\jref{J.~Geophys.~Res.}}
  \def\jqsrt{\jref{J.~Quant.~Spec.~Radiat.~Transf.}}
  \def\memsai{\jref{Mem.~Soc.~Astron.~Italiana}}
  \def\nphysa{\jref{Nucl.~Phys.~A}} \def\physrep{\jref{Phys.~Rep.}}
  \def\physscr{\jref{Phys.~Scr}} \def\planss{\jref{Planet.~Space~Sci.}}
  \def\procspie{\jref{Proc.~SPIE}} \let\astap=\aap \let\apjlett=\apjl
  \let\apjsupp=\apjs \let\applopt=\ao \def\jcap{\jref{JCAP}}
\providecommand{\href}[2]{#2}\begingroup\raggedright\endgroup

\end{document}